\def\puncspace{\ifmmode\,\else{\ifcat.\C{\if.\C\else\if,\C\else\if?\C\else%
\if:\C\else\if;\C\else\if-\C\else\if)\C\else\if/\C\else\if]\C\else\if'\C%
\else\space\fi\fi\fi\fi\fi\fi\fi\fi\fi\fi}%
\else\if\empty\C\else\if\space\C\else\space\fi\fi\fi}\fi}
\def\SP{\let\\=\empty\futurelet\C\puncspace }
\def\etal{et\SP al.\SP }

\def\h-1{$h^{-1}$}
\def\vvec{{\bf v}}
\def\rvec{{\bf r}}

\def\void#1{{}}

\def\h1{$h^{-1}$}

\def\iras{$IRAS$\SP}
\def\datal{da Costa \etal (1998)\SP }
\def\kms{kms$^{-1}$\SP}
\def\etal{et al.\,}
\def\eg{e.g., \,}

\def\lsim{~\rlap{$<$}{\lower 1.0ex\hbox{$\sim$}}}
\def\gsim{~\rlap{$>$}{\lower 1.0ex\hbox{$\sim$}}}
\def\hmu2 {$\bar{\mu}_{1/2}$\SP}

\def\m26 {$m_{26}$\SP}
\def\d26 {$D_{26}$\SP}

\def\mssrs2 {$m_{ssrs2}$\SP}

\def\AJ#1,{AJ, {#1},}
\def\ApJ#1,{ApJ, {#1},}
\def\ApJL#1,{ApJ, {#1},}
\def\ApJS#1,{ApJS, {#1},}
\def\mn#1,{MNRAS, {#1},}
\def\MNRAS#1,{MNRAS, {#1},}
\def\PASP#1,{PASP, {#1},}
\def\Science#1,{Science, {#1},}
\def\Nature#1,{Nature, {#1},}
\def\ZfA#1,{{\sl Zeitschrift f\"ur Astrophysik}, {#1},}

\def\dnsig{$D_n-\sigma$\SP}

\def\puncspace{\ifmmode\,\else{\ifcat.\C{\if.\C\else\if,\C\else\if?\C\else%
\if:\C\else\if;\C\else\if-\C\else\if)\C\else\if/\C\else\if]\C\else\if'\C%
\else\space\fi\fi\fi\fi\fi\fi\fi\fi\fi\fi}%
\else\if\empty\C\else\if\space\C\else\space\fi\fi\fi}\fi}
\def\SP{\let\\=\empty\futurelet\C\puncspace}
\def\iras{{\it IRAS}\SP}
\def\psc{{\it PSCz}\SP}

\def\kms{km~s$^{-1}$\SP}

\def\h1{$h^{-1}$\SP}

\def\etal{{\it et al.\/}\ }

\def\eg{{\it e.g.\/}\rm,\ }
\def\lsim{~\rlap{$<$}{\lower 1.0ex\hbox{$\sim$}}}
\def\gsim{~\rlap{$>$}{\lower 1.0ex\hbox{$\sim$}}}
\def\void#1{{}}

\documentstyle[twocolumn,psfig,rotating]{mn}
\def\puncspace{\ifmmode\,\else{\ifcat.\C{\if.\C\else\if,\C\else\if?\C\else%
\if:\C\else\if;\C\else\if-\C\else\if)\C\else\if/\C\else\if]\C\else\if'\C%
\else\space\fi\fi\fi\fi\fi\fi\fi\fi\fi\fi}%
\else\if\empty\C\else\if\space\C\else\space\fi\fi\fi}\fi}
\def\SP{\let\\=\empty\futurelet\C\puncspace }
\def\etal{et\SP al.\SP }

\def\h-1{$h^{-1}$}
\begin{document}

\title[Comparison of the ENEAR Peculiar Velocities
with the  \psc Gravity Field]
{Comparison of the ENEAR  Peculiar Velocities
with the  \psc Gravity Field}

\author [Nusser et. al.] {Adi Nusser$^1$, Luiz N. da Costa$^{2,3}$,
Enzo Branchini$^4$, Mariangela Bernardi$^5$,\newauthor  M. V. Alonso$^6$, Gary Wegner$^7$, C. N. A. Willmer$^{3,8}$,P. S. Pellegrini$^3$
\\
$^1$Physics Department, Technion- Israel Institute of Technology, 
Haifa 32000, Israel\\
$^2$European Southern Observatory, Karl-Schwarzschild Str. 2, D--85748
Garching b.  M\"unchen, Germany\\
$^3$Observat\'orio Nacional, Rua General Jos\'e Cristino
77, Rio de Janeiro, R. J., 20921, Brazil\\
$^4$ Kapteyn Institute, University of Groningen, Landleven 12, 9700
AV, Groningen, The Netherlands\\
$^5$ 
The University of Chicago, 5640 South Ellis
Avenue, Chicago, IL 60637, USA \\
$^6$ Observatorio Astr\'onomico de
C\'ordoba,  Laprida  854, C\'ordoba, 5000, Argentina\\
$^7$ Dept. of Physics and Astronomy, Dartmouth College, Hanover, 
NH 03755, U.S.A\\
$^8$ UCO/Lick Observatory, University of California, 1156 High Street,
Santa Cruz, CA 95064, U.S.A}

\maketitle

\begin{abstract}
We present a comparison between the peculiar velocity field measured
from the ENEAR all-sky $D_n-\sigma$ catalog and that derived from the
galaxy distribution of the IRAS \psc redshift survey. The analysis is based on a modal
expansion of these data in redshift space by means of spherical
harmonics and Bessel functions. The effective smoothing scale of the
expansion is almost linear with redshift reaching 1500\kms at 3000
\kms.  The general flow patterns in the filtered ENEAR and \psc
velocity fields agree well within 6000\kms, assuming a linear biasing
relation between the mass and the \psc galaxies. The  comparison allows
us to determine the parameter $\beta=\Omega^{0.6}/b$, where $\Omega$
is the cosmological density parameter and $b$ is the linear biasing
factor.  A likelihood analysis of the ENEAR and
\psc modes yields  $\beta=0.5\pm 0.1$, in good agreement with values
obtained from Tully-Fisher surveys.

\end{abstract}


\section {Introduction}

In the standard picture for the formation of cosmic structures via
gravitational instability the peculiar velocity of a galaxy is generated
by fluctuations in the mass distribution. For galaxies outside
virialized systems, linear perturbation theory predicts
\begin{equation}
\vvec (\rvec) \approx {\Omega^{0.6}H_o \over 4\pi} \int{ d^3r^\prime \delta_m
{(\rvec^\prime - \rvec) \over \vert \rvec^\prime - \rvec \vert ^3}} \; ,
\label{lingrav}
\end{equation}
where $\Omega$ is the mass density parameter, $H_o$ is the Hubble
constant and $\delta_m$ is the mass density fluctuation field. If the
relationship between the galaxy distribution, $\delta_g$, and
$\delta_m$ is approximately linear, $\delta_g=b\,\delta_m$, then the
parameter $\beta=\Omega^{0.6}/b$ can be derived from the comparison
between the observed peculiar velocity field and that predicted from
the galaxy distribution.  A particularly useful method for performing
a velocity-velocity comparison is the modal expansion method developed
by Nusser \& Davis (1995, hereafter ND95). This method expands the
velocity fields by means of smooth functions defined in redshift
space, thus alleviating the Malmquist biases inherent in real space
analysis. Furthermore, the modal expansion filters the observed and
predicted velocities in the same way, so that the smoothed fields can
 be compared directly.  Because the number of modes is substantially
smaller than the number of data points, the method also provides the
means of estimating $\beta$ from a likelihood analysis carried out on
a mode-by-mode basis, instead of galaxy-by-galaxy.  The similar
smoothing and the mode-by-mode comparison substantially simplify the
error analysis.  The modal expansion method has previously been used
in comparisons between the 1.2 Jy \iras predicted velocities and
observed velocities inferred from Tully-Fisher (TF) measurements
(Davis, Nusser \& Willick 1996, hereafter DNW, da Costa
et. al. 1998). In this paper, we perform a similar analysis using the
recently completed redshift-distance survey of early-type galaxies
(hereafter ENEAR, da Costa \etal 2000) and the \iras \psc redshift
survey (Saunders et. al. 2000).  Because of differences in the nature
of the data sets considered some slight changes in the method are
required and are described below. Our goal is to investigate how well
the velocity field mapped by early-type galaxies matches the velocity
field inferred from the \psc survey, and to obtain the parameter
$\beta$ yielding the best match.

\void{In the application of the modal expansion method we do not impose
orthogonality of the base functions as in \datal and DNW in their
analysis of TF catalogs. This is because the ENEAR sample
includes individual galaxies and groups of galaxies with different
errors of the distance modulus, making the error matrix of the modes
off-diagonal even if orthogonal base functions are used.}

In section 2, we briefly describe the ENEAR redshift-distance catalog.
In section 3, we describe the modal expansion method as  used here,
present maps of the ENEAR and \psc radial peculiar velocity field and
perform a likelihood analysis to derive $\beta$.  A brief summary of
our conclusions is presented in section 5.

\section {Data}

We use a sub-sample extracted from the all-sky ENEAR redshift-distance
survey (da Costa \etal 2000) comprising 578 objects within $cz \le
6000$~\kms, 355 field galaxies and 223 groups/clusters.  Galaxies have
been objectively assigned to groups and clusters using redshifts taken
from complete redshift surveys sampling the same volume.  Individual
galaxy distances were estimated from an inverse \dnsig template
relation derived by combining cluster data (\eg Bernardi \etal
2000). The cluster sample consists of 569 galaxies in 28 clusters.
Over 80\% of the galaxies in the magnitude-limited sample and roughly
60\% of the cluster galaxies have new spectroscopic and photometric
data obtained by the ENEAR survey. Multiple observations using
different telescope/instrument configurations ensure the homogeneity
of the data. Furthermore, the sample completeness is uniform across
the sky.

\section {The modal expansion}

An unbiased estimate of $\beta = \Omega^{0.6}/b$ can be obtained from
the comparison between smooth velocity fields with similar spatial
resolution, derived from the ENEAR and \psc data.  To generate smooth
fields we expand the peculiar velocities of both data in terms of
smooth base functions.  The expansion carried out here shares the
general properties of that used by ND95, but differs in details.  In
their application to TF catalogs, ND95 defined $P_i\equiv 5 -\log (1 -
u_i/s_i)$, where $s_i=cz_i$ is the galaxy redshift in \kms and $u_i$
its radial peculiar velocity. The function $P$ was then expressed by
an expansion involving smooth functions. The final estimate of the
smoothed velocity field was that obtained by minimizing the scatter of
the rotational speeds given the magnitudes in the inverse TF
relation. The scatter was also simultaneously minimized with respect
to the the parameter of the TF relation.  This led to an unbiased
calibration of the inverse TF relation because the sample was mainly
magnitude selected.  The galaxy angular size and velocity dispersion
in the $D_n-\sigma$ relation do not uniquely fix the magnitude
according to which the ENEAR sample is selected. So simultaneous
minimization might lead to a biased estimate of the parameters of the
$D_n-\sigma$ relation.  Although the bias is mild we use the
calibration of the {\it inverse} $D_n-\sigma$ given by Bernardi~\etal
(2000) by a regression of $\sigma$ on $D_n$ in clusters. We also
express the peculiar velocity, $u$, rather than the function $P$ in
terms of smooth functions.  Another difference is that ND95 used TF
catalogs with all galaxies having the same relative distance error
which allowed an additional simplification in the application of the
modal expansion method, namely, the expansion in terms of orthogonal
smooth functions. This made the TF velocity error covariance matrices
diagonal.  In the ENEAR sample, the relative distance error is not the
same for all objects (galaxies and groups/clusters), so using
orthogonal functions does not offer any further simplification since
the ENEAR error matrix remains non-diagonal. The lack of orthogonality
slightly complicates the error analysis but does not affect the
efficiency of the expansion.  Choosing the spherical harmonics and
Bessel functions to be our base smooth functions we write the radial
peculiar velocity model as
\begin{equation} \tilde u(s,\theta,\phi)\!=\!\! \sum_{l,m,n} \!
\alpha_{nlm}\left[j^{\prime}_{l}\left(k_ny(s)\right)-c_{l1}\right]
Y_{lm}\left(\theta,\phi\right) \quad . \label{pexp} 
\end{equation} where
the sum is over $m=-l$ to $ +l$, $l\!=\!0$ to $ l_{max}$, and
$n\!=\!0$ to $n_{max}$.  For the reasons given in DNW, we formulate
our model to describe the velocity field with respect to the motion of
the Local Group.  The constant $c_{l1}$ is non-zero only for the
dipole term ensuring that $u=0$ at the origin.  The function $y(s)$ in
the argument of the Bessel functions makes their oscillations match
the radial distribution of the ENEAR data. Here we take $ y^2=
\ln[1+(\frac{s}{1000})^2]$, but other similar forms can be used as
well.  Also the expansion does not include a Hubble-like flow
($u\propto s$) so we assume that any such flow has been consistently
removed from the ENEAR and \psc velocities.  The coefficients
$\alpha_{nlm}$ are found by minimizing
\begin{equation}
\label{minchi} \chi^2 = 
\sum  \sigma^{-2}_i [ \tilde u_i - u^{\mathrm o}_i]^2 
\end{equation}
where $u^o_i$ are the raw observed velocities and $\sigma_i$ is the
error of the velocity estimate resulting from observational
uncertainties and intrinsic scatter in the $D_n-\sigma$.  For field
galaxies $\sigma_i =0.23 s_i$ and for groups of galaxies it is 
reduced by $1/\sqrt{N_g}$, where $N_g $ is the number of galaxies in
the group.

\section {Smooth  Velocity Maps and the determination of $\beta$}

We apply the modal expansion method to smooth the raw measured
velocities of the $578$ ENEAR objects within a redshift of 6000~\kms
(Bernardi et. al. 2000).  We use 51 modes corresponding to
$l_{max}=4,n_{max}=3$ in (\ref{pexp}). The smoothing scale of these
functions is linear with redshift and matches the low resolution
filter used in \datal (see their Figure 1).  The smoothed velocities
were then derived by minimizing (3) with respect to $\alpha_{lmn}$
assuming an error of $\sigma_{i}=0.23s_i/\sqrt{N_g}$ in the raw
velocities of the ENEAR objects. The reduced $\chi^2$ per d.o.f of the
fit was 1.017, a satisfactory value in this type of analysis (see DNW,
da Costa et al. 1998).

Given an assumed value for $\beta$ we interpolate the \psc predicted
velocity field, computed by Branchini et. al. (2000), to the positions
of the ENEAR galaxies.  Branchini et. al. obtained the \psc velocities
from the \psc galaxy distribution with a Top-Hat window of width
equal to half the mean particle separation at a given redshift.  The
\psc fields are then expanded in the same orthogonal set of basis
functions as employed for the ENEAR velocities.  The \psc and ENEAR
velocities are guaranteed to have the same resolution because the
original smoothing of the \psc density field is small compared to the
resolution of the modal expansion.

\begin{figure}
\vspace{1cm}
\centering
\mbox{\psfig{figure=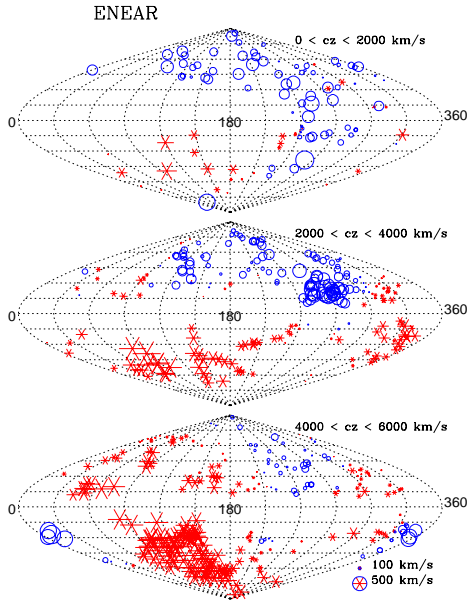,height=8cm,bbllx=18pt,bblly=144pt,bburx=592pt,bbury=718pt}}
\caption{The sky projection in galactic coordinates  as seen in the
 LG frame of the filtered ENEAR velocity field.}
\label{fig:filt}
\end{figure}

\begin{figure}
\vspace{1cm}
\centering
\mbox{\psfig{figure=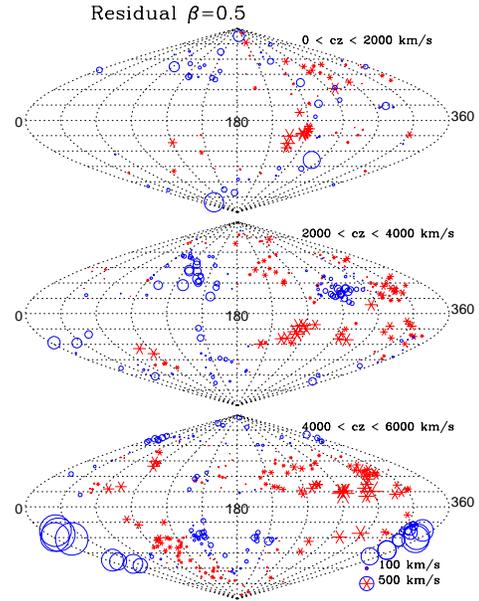,height=8cm,bbllx=18pt,bblly=144pt,bburx=592pt,bbury=718pt}}
\caption{The  residual velocity field (ENEAR minus PSCz)  
for $\beta =0.5$ }
\label{fig:rez05}
\end{figure}

The smoothed ENEAR velocities are shown in Figure~\ref{fig:filt}, in
redshift shells 2000\kms thick. Comparison of this figure and Figure~3
of \datal shows that the general flow pattern is remarkably
similar. In the case of ENEAR, in the innermost shell very few
prominent structures are probed by bright ellipticals. However, in the
next two shells a strong dipole pattern can be easily recognized,
being of comparable amplitude to that of observed with the SFI
galaxies. This dipole corresponds to the reflex motion of the Local
Group, with infalling galaxies in the Hydra-Centaurus direction and an
outflow towards the Perseus-Pisces complex.\void{ For comparison, we
show in Figure~\ref{fig:psc05} the
\psc field reconstructed with $\beta=0.5$ and filtered with the same
basis functions as ENEAR. This value for $\beta$ was chosen because as
shown below it yields the best match between the ENEAR and the \psc
smoothed velocity fields.} The quality of the match can be evaluated
from Figure~\ref{fig:rez05} which shows the residual velocity field
obtained subtracting the smoothed \psc field from that of the ENEAR,
assuming $\beta=0.5$. As can be seen the overall agreement is good
with only a few more distant galaxies giving large residuals. Note,
however, that even though with a larger amplitude, the mismatch seen
in the outermost redshift shell at $l \sim 0^\circ$, $-60^\circ \lsim b
\lsim -15^\circ$ between ENEAR and \psc correspond to mismatches in the
comparison between SFI and 1.2 Jy \iras velocity fields.  This may
correspond to a real mismatch between measured and predicted
velocities which deserves further investigation.

The filtered ENEAR and \psc velocity fields are fully described by the
modal expansion coefficients, $\alpha_{en}$ and $\alpha_{ps}$, of the 
ENEAR and \psc fields, respectively.
Since the number of these coefficients is significantly smaller than
the number of galaxies, it is more efficient to estimate $\beta$ by
comparing the modes rather than the individual galaxy velocities.  As in
\datal we define our best estimate for $\beta$ as the value that corresponds
to the minimum of the pseudo-$\chi^2$
\begin{equation}
\tilde \chi^2(\beta)=\sum_{j, j^\prime}\left[\alpha^j_{en}-\alpha^j_{ps}(\beta)\right]
\left[{\rm \bf T}+ {\rm \bf M}(\beta)\right]^{-1}
\left[\alpha^{ j^\prime}_{en}-\alpha^{j^\prime}_{ps}(\beta)\right] ,
\label{chia}
\end{equation}
where ${\rm \bf T\equiv}<\delta \alpha^j_{en} \delta
\alpha^{j^\prime}_{en}>$ 
and ${\rm \bf M}\equiv <\delta \alpha^j_{ps} \delta
\alpha^{j^\prime}_{ps}>$ are the the error covariance matrices of
the coefficients $\alpha^j_{en}$ and $\alpha^j_{ps}$, respectively.
For brevity of notation we have replaced the triplet $n,l,m$ with one
index $j$.  The \psc covariance matrix ${\rm \bf M}$ incorporates
errors due to (i) the uncertainty in the LG motion, which creates a
dipole discrepancy between the ENEAR and the \psc velocities, $(ii)$
the discreteness in distribution of galaxies which propagates into the
velocity field. and $(iii)$ small scale coherent (as in triple valued
zones) nonlinear velocities that are not included in the \psc
recovered velocities.  Details of how these error contributions are
computed are in \datal. Since the expansion functions are
not orthogonal the ENEAR covariance matrix $\rm \bf T$ has
nonzero off-diagonal elements. This matrix is simply the inverse of
$\frac {\partial^2 \chi^2} {\partial \alpha_{en}^j\partial
\alpha_{en}^{j^\prime}}$ where the derivatives are computed at the
minimum of $\chi^2$ given by (\ref{minchi}).

\begin{figure}
\vspace{1cm}
\centering
\mbox{\psfig{figure=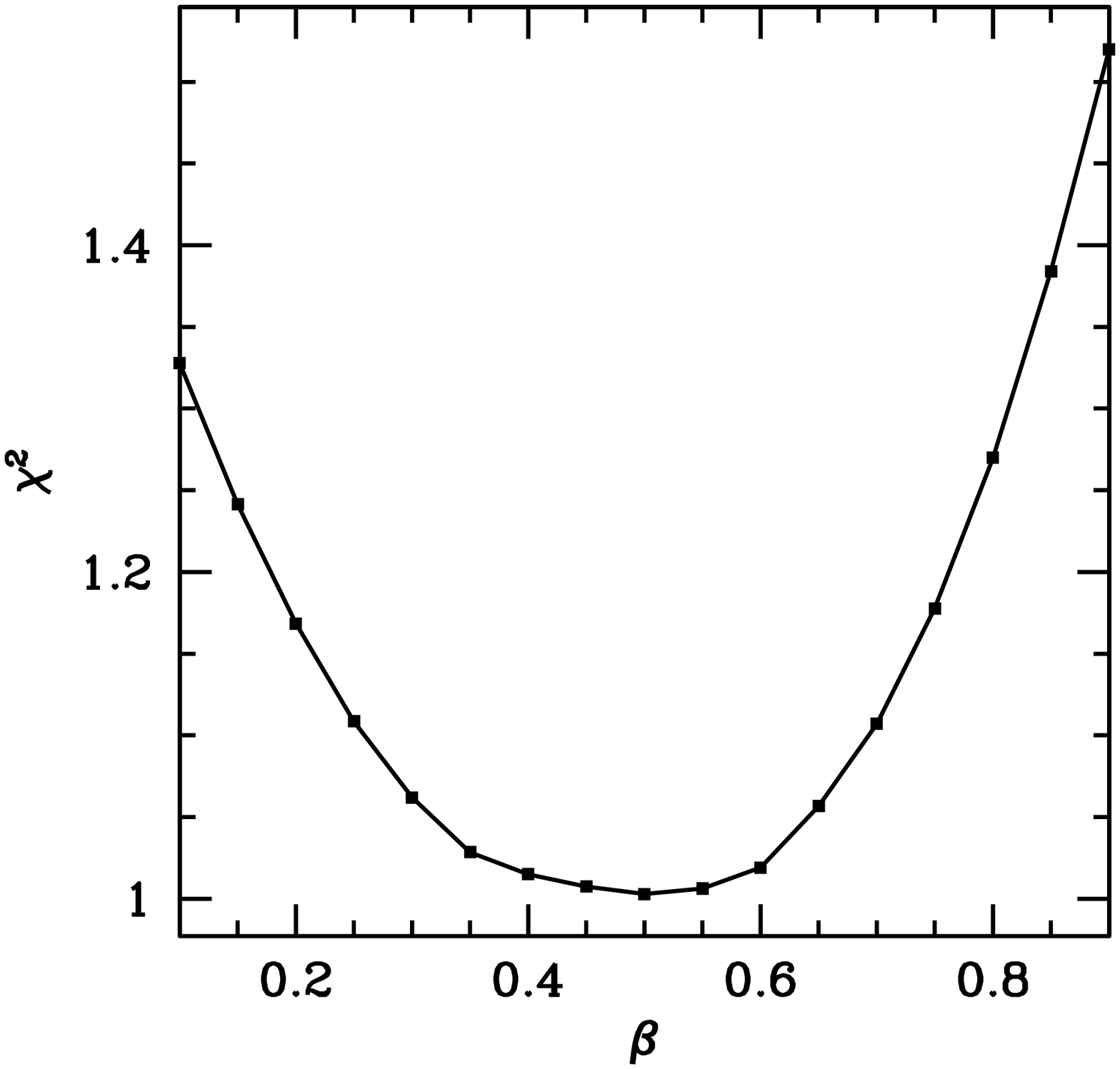,height=8cm,bbllx=18pt,bblly=144pt,bburx=592pt,bbury=718pt}}
\mbox{\psfig{figure=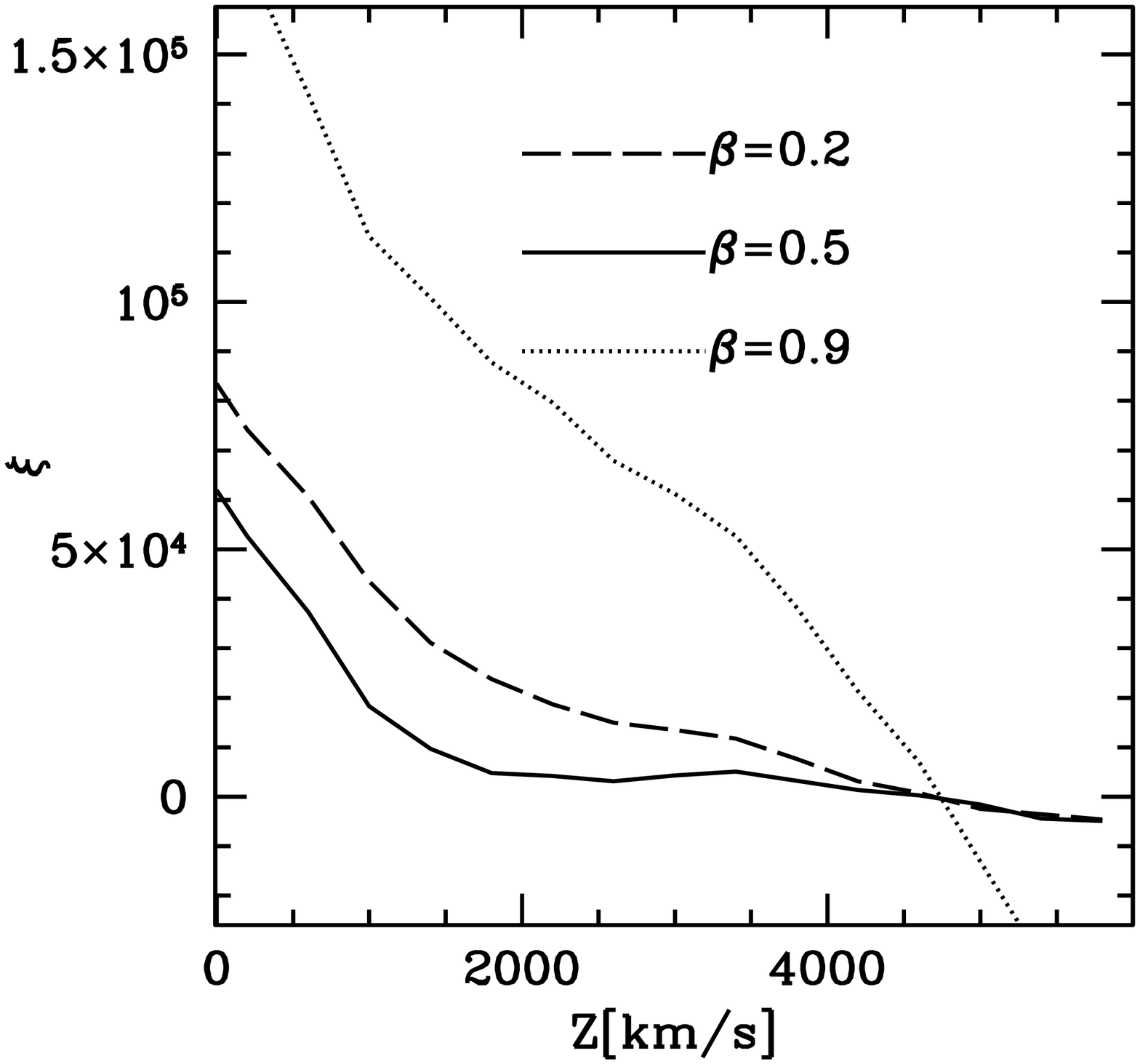,height=8cm,bbllx=18pt,bblly=144pt,bburx=592pt,bbury=718pt}}
\caption{{\it Top panel}: 
curve of reduced pseudo-$\chi^2$ versus $\beta$ computed
using equation (5). {\it Bottom panel}:  the correlation function of the
velocity residual field for $\beta=0.2$, $0.5$ and $ 0.9$.}
\label{fig:dv}
\end{figure}

Given the covariance matrices, we compute the curve of the reduced
$\tilde \chi^2(\beta)$ as a function of $\beta$, which is shown in the
top panel of Figure~\ref{fig:dv}.  The curve was computed with an
error of 150~\kms in the estimation of the LG motion and 160~\kms for
the amplitude of nonlinear error in the \psc field (see da Costa
et. al. 1998).  This amplitude of the nonlinear error was chosen to
make the $\tilde {\chi}^2$ per d.o.f equal to unity at the minimum. In
their analysis of the SFI and 1.2 Jy \iras, \datal obtained a lower
value of 90~\kms for the amplitude of this error.  The difference can
be attributed, as expected, to a better match between the SFI and
\iras velocities and the increased nonlinearities in the \psc
velocity at the positions of the ENEAR galaxies which preferentially
reside in high density regions.

The minimum value of the $\tilde \chi^2$ is attained at $\beta=0.5$,
with the 1-sigma error being less than $0.1$. We note that this result
is not sensitive to the exact values adopted for the error estimates
% and found that the location of the minimum remained at $\beta= 0.5\pm 1 $
%for all reasonable estimates of the errors. It is interesting to
%inspect the correlation function of the residual fields.
Another statistic indicating the goodness of the match between the
fields for various $\beta$ is the correlation function of the residual
$u_{en}-u_{ps}$ between the smoothed ENEAR and \psc radial velocities.
This is shown in the bottom panel of Figure~\ref{fig:dv} for
$\beta=0.2$, 0.5, and 0.9.  The amplitude of the \psc field is small
for $\beta=0.2$, so the correlation function for this $\beta$ is close
to the correlation function of $u_{en}$ alone, while the opposite is
true for $\beta=0.9$.  On the other hand, for $\beta=0.5$ the
correlation of the residual velocity field is significantly smaller,
indicating a good match between the measured and predicted velocity
fields.

\section{Summary and Discussion}

Using the modal expansion method of ND95 and the recently completed
ENEAR redshift-distance survey and the \psc redshift survey we have
carried out a comparison between the observed peculiar velocity field
and that predicted from the distribution of \psc galaxies.  We find
that the corresponding smoothed fields agree well and the best match
is obtained with $\beta=0.5 \pm 0.1$.  This value is intermediate to
those derived using the Mark~III and SFI catalogs both based primarily
on spiral galaxies.  It is also consistent with the results obtained
by Borgani \etal (2000) using an independent method based on modeling
the velocity correlation function. Note, however, that the discrepancy
between the values determined from these methods and those obtained
from the power spectrum analysis (\eg Zaroubi \etal 2000) and
density-density comparisons (\eg Sigad \etal 1998) still persist.  
The good agreement between
SFI and 1.2 Jy \iras and between ENEAR and \psc implies that the SFI
and ENEAR velocity fields are also in good agreement. This suggests
that the velocity maps obtained from the new distance-redshift surveys
are a fair representation of the underlying velocity field, as the
general characteristics of the observed flow fields are independent of
the type of galaxies and distance indicators used.  The good agreement
among the values of $\beta$ obtained using Mark~III, SFI, ENEAR, 1.2
Jy and
\psc catalogs gives further support to low values of $\beta$ and point
toward low-density cosmologies.

\section*{Acknowledgements}

The authors would like to thank M. Maia, C. Rit\'e and O. Chaves for
their contribution over the years. The results of this paper are based
on observations conducted at the European Southern Observatory (ESO)
and the MDM Observatory. CNAW acknowledges support from NSF AST95-29028.

\end{document}